\documentclass{article}%
\usepackage{amsmath}
\usepackage{amsfonts}
\usepackage{amssymb}
\usepackage{graphicx}%
\newcommand{\bpartial}{\mathop{\partial\kern -4pt\raisebox{.8pt}{$|$}}}
\newcommand{\bra}{\mathopen{[\kern-1.6pt[}}
\newcommand{\ket}{\mathclose{]\kern-1.5pt]}}
\newcommand{\bbra}{\mathopen{[\kern-2.2pt[\kern-2.3pt[}}
\newcommand{\bket}{\mathclose{]\kern-2.1pt]\kern-2.3pt]}}

\makeindex
\begin{document}

\title {\large{ \bf Two dimensional Nambu sigma model}}

\vspace{3mm}

\author {  \small{ \bf S. Farhang-Sardroodi }\hspace{-2mm}{ \footnote{ e-mail: m.farhang88@azaruniv.edu}} { \small
and} \small{ \bf A. Rezaei-Aghdam }\hspace{-2mm}{
\footnote{Corresponding author. e-mail:
rezaei-a@azaruniv.edu}} \\
{\small{\em Department of Physics, Faculty of Science, Azarbaijan Shahid Madani University, }}\\
{\small{\em   53714-161, Tabriz, Iran  }}}

\maketitle

\begin{abstract}
We present two dimensional sigma model by using the Nambu structure
on a manifold in general, and on a Lie group as a special case.
Then, we consider model constructed from Nambu structure of order
three and obtain conditions under which this model is equivalent to
a WZW model. Furthermore, we present an example for this case on the
four dimensional Heisenberg Lie group. Finally, as another example
we show that the model constructed with Nambu structure of order
three on the central extension of the 2D Poincare Lie group is
integrable.
\end{abstract}
\section{\bf Introduction}

Two dimensional sigma models have important role in string theories
\cite{Pol}-\cite{Cal1}, statistical and quantum mechanical solvable
models \cite{Fri},\cite{Alv}. Their symmetries are deeply depend on
the geometry of their target spaces. The metric and antisymmetric
tensor fields in the sigma model in general have no special
structures. Of course up to now there are some attempts to use the
special geometric structures in the model. For example, in the
Poisson sigma model \cite{SS} the Poisson structure is used in the
model. The advantage of this model is that it contains other two
dimensional field theories (such as two dimensional Yang-Mill, two
dimensional gravity, BF model, topological sigma model and gauged
WZW models) as special cases. Another example is the Hitchin sigma
model \cite{Zuc} such that in these models the generalized complex
geometry conditions are obtained from a master equation of the
master action (sigma model) \cite{Bat} containing the generalized
geometric structures. Our main idea is the construction of different
sigma models from different geometric structures on the manifolds
\cite{Rez1}. In this direction, in our previous work \cite{Rez2} we
presented \emph{Nijenhuis sigma model}. Here, we present \emph{Nambu
sigma
model} constructed from Nambu structure on the manifold.\\
The paper is organized as follows. In section two we review some
basic definitions and properties of the Nambu structure. Then, in
section three by use of the Nambu structures of any order (even or
odd) and an antisymmetric tensor of second rank, we present Nambu
sigma models; specially we focus on the models constructed by Nambu
structure of order three and investigate these models on Lie groups.
After obtaining the conditions under which the Nambu sigma model is
equivalent to the WZW model; in section four we present two
examples: the Nambu sigma models on the Heisenberg and 2D extantion
of Poincare Lie group. In the former we show that the model is
equaivalent to the WZW models and then in section five we show that
the second model is an integrable model.
\section{\bf Some basic definitions}
In this section for self containing of the paper we review some
basic definitions on Nambu structure. In 1973 Nambu \cite{Nam}
studied a dynamical system which was defined as a Hamiltonian system
with respect to Poisson-like bracket, defined by a Jacobian
determinant. About two decades later Takhtajan \cite{Tak} by using
an axiomatic formulation for $n$-bracket introduced the concept of
Nambu structure and gave the basic properties and geometric
formulations of Nambu manifolds. This new approach motivated a
series of papers about some new concepts in relation to Nambu
structure; also another generalization was the so-called generalized
Poisson bracket \cite{Azc1}; (a comparison of both concepts was
given in\cite{Azc2}). A $ C^{\infty}$ manifold endowed with a Nambu
tensor (a skew-symmetric contravariant tensor field) is called Nambu
manifold if the induced bracket satisfies the fundamental identity,
which is generalization of the usual Jacobi identity
\cite{Gau}-\cite{Nak1}. Let $M$ be a smooth $n$-dimensional manifold
and $ C^{\infty}(M)$
 denotes the algebra of differentiable real-valued functions on $M$. A
 Nambu structure of order $m$ is given by an m-dimensional
 multivector field, i.e. a $ C^{\infty}(M)$-skew multilinear map
\begin{eqnarray}
\eta:\Omega(M)\times...\times\Omega(M)\longrightarrow
{C^{\infty}(M)},\label{2.1}
\end{eqnarray}
where $\Omega(M)$ is the set of one forms on $M$; such that in terms
of the local coordinates $(x^1,x^2,...x^n)$ we have
\begin{eqnarray}
\eta=\eta^{\mu_{i_1}...\mu_{i_m}}(x)\frac{\partial}{\partial
x^{\mu_{i_1}}}\wedge\frac{\partial}{\partial
x^{\mu_{i_2}}}...\wedge\frac{\partial}{\partial
x^{\mu_{i_m}}},\label{2.2}
\end{eqnarray}
where summation over repeated indices is understood. In this manner
one can define the bracket of $m$ functions $f_1,...,f_m\in
\cal{F(M)}$ as follows:
\begin{eqnarray}
\{f_1,f_2,...,f_m\}=\eta(df_1,df_2,...,df_m),\label{2.3}
\end{eqnarray}
 note that the left hand side of the above equation is the
 Nambu bracket of $m$ functions which is the generalize action of the Poisson
 bracket. Furthermore, since the bracket satisfies Leibnitz rule, one can
define a vector field $X_{f_1,...,f_{n-1}}$by
\begin{eqnarray}
X_{f_1,...,f_{n-1}}(\textit{g})=\{f_1,...,f_{n-1},\textit{g}\}\;\;\;\forall{f_1,...,f_{n-1},g
}\in \cal{F(M)},\label{2.4}
\end{eqnarray}
where this vector field is called \emph{Hamiltonian }vector field; the space of \emph{Hamiltonian} vector field is denoted by $\cal{H}$.\\
\textbf{Definition 1}:\cite{Nak1}-\cite{Nak3}\emph{An element
$\eta\in \Gamma(\Lambda^mTM) $, for $m\geq 3$ is called a Nambu
tensor of order $m$ if it satisfies ${\cal L}_{X}\eta=0$, for all
}$X\in \cal{H} $; \emph{where ${\cal L}$ stands for Lie
derivative.}\\ The above definition is clearly equivalent to the
following fundamental identity \cite{Tak}
\begin{eqnarray}
 \{f_1,...,f_{n-1},\{g_1,...,g_n \}\}=\{\{f_1,...,f_{n-1},g_1 \},g_2...,g_n \} + \{g_1,\{f_1,...,f_{n-1},g_2 \},g_3...,g_n \}\nonumber\\
 +...+ \{g_1,...g_{n-1},\{f_1,...,f_{n-1},g_n
 \}\},\label{2.5}~~~~~~~~~~~~~~~~~~~~~~~~~~~~~~~~~
\end{eqnarray}
 for all $f_1,...,f_{n-1},g_1,...,g_n \in{ C^{\infty}(M)}.$ If $n=2$, this
 equation is nothing but the Jacobi identity for the Poisson structures.
 In Ref. \cite{Far}, we have found the Nambu structures of order three and
 four on real four dimensional Lie groups. Now, in the next section
 by use of the Nambu structures of any order in general, and as an example in the especial case of order three we present a two
 dimensional sigma model (\emph{Nambu sigma model}) on a manifold in general and on a Lie group
 as a special case. As an example, some sigma models constructed from Nambu
 structure of order $m=3$ are also considered.
\section{\bf Nambu-Sigma model }
Now, we assume that the manifold $M$ has the Nambu structure up to
top order $n$ ( $\eta^{\mu_1,...,\mu_n}$), a 2-form $w_{\mu\nu}$ and
also a 3-form $H_{\mu\nu\lambda}$; then one can construct the
following actions by using of the Nambu structure with even order
and odd order, respectively\footnote{Here ``~$\hat{}$~" stands for
omitted term.}
\begin{eqnarray} S_1=\sum_{j,l=1}^{2k}\int_{\Sigma} d^2\sigma
\varepsilon^{\alpha\beta}\eta^{\mu_{i_1}\mu_{i_2}...\mu_{{i_{2k}}}}w_{\mu_{i_1}\mu_{i_2}}...\hat{w}_{\mu_{i_j}\mu_{i_l}}...w_{\mu_{i_{2k-1}}\mu_{i_{2k}}}
G_{\mu_{i_{j}}\lambda}G_{\mu_{i_{l}}\nu}\partial_{\alpha}
x^{\lambda}\partial_{\beta}x^{\nu},\label{3.1}
\end{eqnarray}
\begin{eqnarray} S_2=\sum_{j,l=1}^{2k+1}\int_{B} d^3\sigma
\varepsilon^{\alpha\beta\gamma}\eta^{\mu_{i_1}\mu_{i_2}...\mu_{{i_{2k+1}}}}H_{\mu_{i_1}\mu_{i_2}\mu_{i_3}}...\hat{H}_{\mu_{i_j}\mu_{i_l}\mu_{i_p}}...
H_{\mu_{i_{2k-1}}\mu_{i_{2k}}\mu_{i_{2k+1}}}~~~~~~~~~
\nonumber\\\times
G_{\mu_{i_{j}}\rho}G_{\mu_{i_{l}}\nu}G_{\mu_{i_{p}}\lambda}
\partial_{\alpha}
x^{\rho}\partial_{\beta}x^{\nu}\partial_{\gamma}x^{\lambda},\label{3.2}
\end{eqnarray}
where B is a three-dimensional manifold with boundary $\partial
{B}=\Sigma$, and the coordinates $x^{\mu}$ are maps from $ \Sigma$
to $M $(such that those can be extend in an arbitrary fashion to the
maps from $B$ to $M$). In this way, by adding these actions to the
following chiral action:
\begin{eqnarray}
S_{ch}=\int_{\Sigma}d^2\sigma G_{\mu\nu}\partial_{\alpha}
x^{\mu}\partial^{\alpha} x^{\nu},\label{3.3}
\end{eqnarray}
we construct new two dimensional sigma models, which we call two
dimensional \emph{Nambu sigma models}.\footnote{Note that these
sigma models, are different from the models presented in
\cite{Jurco:2012yv}.} Note that these models in general are not
conformal or integrable, and
one can obtain conditions on $\eta$, $w$ and $H$ under which these models have the above properties. \\
Here, among these models we focus on the following model constructed
only by Nambu structure of order three
\begin{eqnarray}
S_{Ns}=\int_{\Sigma}d^2\sigma G_{\mu\nu}\partial_{\alpha}
x^{\mu}\partial^{\alpha} x^{\nu} +k\int_Bd^3\sigma \epsilon_{\alpha
\beta
\gamma}\eta^{\mu\nu\lambda}G_{\mu\eta}G_{\nu\rho}G_{\lambda\sigma}\partial^{\alpha}
x^{\eta}\partial^{\beta} x^{\rho}\partial^{\gamma}
x^{\sigma}.\label{3.4}
\end{eqnarray}
Note that for this cases the forms $w$ and $H$ have disappeared and
also $k$ is a parameter. In this general form, this new model is not
conformal or integrable; but one can find conditions by imposing
conformality or integrability to the model. Here, we focus on the
model \eqref{3.4} over a Lie group. The Nambu sigma model on a Lie
group G can be written as follows: {\small\begin{eqnarray}
S_{Ns}=\int_{\Sigma}\Omega_{ij}{(g^{-1}dg)^{i}}\wedge{(g^{-1}dg)^{j}}
+k\int_B\eta^{ijk}\Omega_{il}\Omega_{jm}\Omega_{kn}{(g^{-1}dg)^{l}}\wedge
{(g^{-1}dg)^{m}}\wedge{(g^{-1}dg)^{n}},\label{3.5}
\end{eqnarray}}
$\forall g \in G$, such that with the assumption $\{X_i\}$ as a
basis of Lie algebra \textbf{g} of the Lie group G we have
$g^{-1}dg=(g^{-1}dg)^iX_i=e^{i}_{~\mu}dx^{\mu}$. Here, $
e^{i}_{~\mu}$ is vierbein such that the metric and Nambu structure
of order three on the Lie group G can be written as
$G_{\mu\nu}=e^i_{~\mu}\Omega_{ij}e^{~j}_\nu,\;\eta^{\mu\nu\lambda}=\eta^{ijk}e_i^{~\mu}
e_j^{~\nu} e_{~k}^\lambda$ where $\Omega_{ij}$ and $\eta^{ijk}$ are
the metric and Nambu structure on the Lie algebra \textbf{g}, and
$e_i^{~\mu}$ is the inverse of $e^i_{~\mu}$. Note that when
$\eta^{\mu\nu\lambda}$ is a constant or is a linear function of the
 coordinates $x^\mu$, then our model (\eqref{3.4} or \eqref{3.5}) in general is not conformal or
integrable. Here we obtain condition on the algebraic Nambu
structure $\eta^{ijk}$ such that the second term in the model
\eqref{3.5} is the same as the WZW term. We know that the WZW term
can be written as follows:{\small \begin{eqnarray}
I_{wzw}=\int_{B}<(g^{-1}dg)\wedge[g^{-1}dg\wedge
g^{-1}dg]>=\int_{B}(g^{-1}dg)^l\wedge(g^{-1}dg)^m\wedge
(g^{-1}dg)^nf^i_{~mn}\Omega_{li}.\label{3.6}
\end{eqnarray}}
 Now by comparing this term with the second term of \eqref{3.5} we will obtain the following relation between the Nambu
 structure $\eta^{ijk}$
  and structure constant $f^i_{~jk}$ on the Lie algebra
\begin{eqnarray}
f^i_{mn}=\eta^{ijk}\Omega_{jm}\Omega_{kn},\label{3.7}
\end{eqnarray}
 or in the matrix form
 \begin{eqnarray}
{\cal Y}^i=-\Omega\eta^{i}\Omega,~~\label{3.8}
\end{eqnarray}
where $({\cal Y}^{i})_{mn}=-f^i_{mn}$ is the adjoint representation
of the Lie algebra \textbf{g} and $(\eta^{i})^{kj}=\eta^{ikj}$. Note
that this relation is compatible with the ad-invariant metric
$\Omega$ (the metric satisfies the relation
$\chi_i\Omega=-(\chi_i\Omega)^t$ with
$(\chi_i)_m^{~n}=-f_{im}^{n}$).
 In this way, we see that for some special examples, where \eqref{3.8} holds, the Nambu sigma model is reduced to the WZW model as a special
 case. In other words according to the C theorem \cite{Z} this Nambu
 sigma model with the above property is a fixed point of the RG flow
 of the general Nambu sigma model with Nambu structure of order
 three. Now, in the following we present an example for this case and also
other example for the model \eqref{3.5} on the Lie group which is
integrable.
\section{Examples}

In this section we consider two examples for the Nambu sigma model
on a Lie group with Nambu structure of order three (model
 \eqref{3.5}).\\

 \textbf{a}) We first consider the four dimensional Heisenberg Lie group.
 This Lie group has Lie algebra which is isomorphic with the \textbf{$A_{4,8}$} in the Petra and \emph{et al.} \cite{PSWZ} classification of
 four dimensional real Lie algebras. We have the following commutation relations for this algebra
 \begin{eqnarray}
[P_2,T]=P_2,\;\;\;\;[P_2,J]= P_1,\;\;\;\;[T,J]=J,\label{4.1}
\end{eqnarray}
where we use the generators ${X_i}=\{P_1,P_2,J,T \}$. Recently in
\cite{Far}, we have obtained the Nambu structures of order four
 and three for its related Lie  sub group; such that a Nambu structure of
order three has the following form
\begin{eqnarray}
\eta=(q_{2}a_{2}+q_{3}u)\frac{\partial}{\partial
a_1}\wedge\frac{\partial}{\partial
a_2}\wedge\frac{\partial}{\partial u}.\label{4.2}
\end{eqnarray}
 By use of the following parametrization for the Lie group elements
 $g\in A_{4,8}$
 \begin{eqnarray}
g=e^{\Sigma_{i}a_{i}P_{i}}e^{uJ}e^{vT},~~~~\label{4.3}
\end{eqnarray}
 and vierbein \cite{Far}
\begin{eqnarray}
e^i_{~\mu}=k\left(
              \begin{array}{cccc}
                1 & u& 0 & 0 \\
                0 & e^v & 0 & 0 \\
                0 & 0 & e^{-v} & 0 \\
                0 & 0 & 0 & 1\\
              \end{array}
              \right)
,\label{4.4}\end{eqnarray}and after some calculations we obtain
\begin{eqnarray}
\eta^{ijk}=-1.~~~~~~~~~~~~~~~~~~~\label{4.5}
\end{eqnarray}
We see that for this example the relation (\eqref{3.7} or
\eqref{3.8}) holds with the following ad-invariant metric for
\textbf{$A_{4,8}$} \cite{Far}
\begin{eqnarray}
\Omega_{ij}=\left(
              \begin{array}{cccc}
                0 & 0 & 0 & -k \\
                0 & 0 & k & 0 \\
                0 & k& 0 & 0 \\
                -k & 0 & 0 & b\\
              \end{array}
              \right)
.\label{4.6}\end{eqnarray} In other words, the Nambu model
\eqref{3.5} on Lie
group  \textbf{$A_{4,8}$} is reduced to the WZW model.\\

 \textbf{b})The second model we will consider in the following is the Nambu sigma model based on a certain non-semi-simple
 Lie algebra of dimension four. This Lie algebra has the following
 explicit description:
\begin{eqnarray}
[J,P_i]=\varepsilon_{ij}P_i,\;\;\;\;[P_i,P_j]=
\varepsilon_{ij}T,\;\;\;\;[T,J]=[T,P_a]=0.\label{4.7}
\end{eqnarray}
Indeed this Lie algebra is a central extension of the 2D
Poincar$\acute{e}$ algebra which is reduced to Euclidian algebra,
when one sets $T=0$. The corresponding simply-connected Lie group is
called R \cite{Edw}\footnote{Note that the Lie algebra \eqref{4.7}
is
isomorphic with the Lie algebra $A_{4,10}$ of \cite{PSWZ}.}.\\
For writing the action of the Nambu sigma model \eqref{3.5} we use
the following parametrization of the Lie group manifold of the above
Lie algebra \cite{Edw}
\begin{eqnarray}
g=e^{\Sigma_{i}a_{i}P_{i}}e^{uJ+vT}.\label{4.8}
\end{eqnarray}
In this way, the left invariant 1-forms and vector fields are
obtained as follows\cite{Edw}
\begin{eqnarray}
g^{-1}dg=(cosu\;da_k+\varepsilon_{jk}sinu\;da_k)P_{k}+d_kuJ+(dv+\frac{1}{2}\varepsilon_{jk}a_kda_j)T,\label{4.9}\\
X_{1}=cosu\frac{\partial}{\partial a_1}+sinu\frac{\partial}{\partial
a_2}-a_{2}cosu\frac{\partial}{\partial v},\label{4.10}\\
X_{2}=-sinu\frac{\partial}{\partial
a_1}+cosu\frac{\partial}{\partial
a_2}+a_{2}sinu\frac{\partial}{\partial v},\label{4.11}\\
X_{3}=\frac{\partial}{\partial u}\;\; X_{4}=\frac{\partial}{\partial
v}.\label{4.12}
\end{eqnarray}
In \cite{Far}, we have obtained the Nambu structure of order four
and three for this Lie group; where the Nambu structure of order
three has the following form
\begin{eqnarray}
\eta=(q_{1}a_{1}+q_{2}a_{2})\frac{\partial}{\partial
a_1}\wedge\frac{\partial}{\partial
a_2}\wedge\frac{\partial}{\partial v}.\label{4.13}
\end{eqnarray}
The Nambu structure of order three on the extension of the 2D
Poincare can also be written as follows \cite{Far}:\footnote{Indeed
this is the Nambu structure of top order for the Lie subalgebra of
2D Poincare algebra with basis $A_{3,1}:\{X_1,X_2,X_4
\}$\cite{Far}.}
\begin{eqnarray}
J=\frac{\partial}{\partial a_1}\wedge\frac{\partial}{\partial
a_2}\wedge\frac{\partial}{\partial v},\label{4.14}
\end{eqnarray}
such that $\eta^{124}=1$. Now using \eqref{4.9} and the following
Lie algebra metric $\Omega_{ij}$ \cite{Edw}
\begin{eqnarray}
\Omega_{ij}=k\left(
              \begin{array}{cccc}
                1 & 0 & 0 & 0 \\
                0 & 1 & 0 & 0 \\
                0 & 0 & b & 1 \\
                0 & 0 & 1 & 0 \\
              \end{array}
              \right)
,\label{4.15}\end{eqnarray} one can obtain the first term of the
action \eqref{3.5} as the first term of \eqref{3.4} with the
following metric \cite{Edw}
\begin{eqnarray}
G_{\eta\mu}=\left(
              \begin{array}{cccc}
                1 & 0 & \frac{a_2}{2} & 0 \\
                0 & 1 & \frac{-a_1}{2} & 0 \\
                \frac{a_2}{2} & \frac{-a_1}{2} & b & 1 \\
                0 & 0 & 1 & 0 \\
              \end{array}
              \right)
.\label{4.16}\end{eqnarray} In the same way, using the above Nambu
structure \eqref{4.14} and the Lie algebra metric $\Omega_{ij}$, the
second term of Nambu sigma model action \eqref{3.5} can be
integrated as follows:
\begin{eqnarray}
\int_{\Sigma}d^2\sigma
2\varepsilon_{\alpha\beta}(a_{1}\partial^{\alpha}a_{2}\partial^{\beta}u+u\partial^{\alpha}
a_{1}\partial^{\beta}a_{2}+a_{2}\partial^{\alpha}u\partial^{\beta}a_{1}),\label{4.17}
\end{eqnarray}
such that by comparison of the above term with the second term of
the following action
\begin{eqnarray}
S_{Ns}=\int_{\Sigma}d^2\sigma (G_{\mu\nu}\partial_{\alpha}
x^{\mu}\partial^{\alpha}
x^{\nu}+B_{\mu\nu}\varepsilon_{\alpha\beta}\partial^{\alpha}x^{\mu}\partial^{\beta}x^{\nu}),\label{4.18}
\end{eqnarray}
we have
\begin{eqnarray}
B_{\mu\nu}=\left(
              \begin{array}{cccc}
                0 & u & -a_{2} & 0 \\
                -u & 0 & a_{1}& 0 \\
                a_{2} & a_{1} & 0 & 0\\
                0 & 0 & 0 & 0 \\
              \end{array}
              \right)
,\label{4.19}\end{eqnarray} where in this action
$x^{\mu}=(a_{1},a_{2},u,v)$. By comparing this action with the WZW
action on R \cite{Edw}, i.e.,
\begin{eqnarray}
S_{wzw}(g)=\int_{\Sigma}d^2\sigma (G_{\mu\nu}\partial_{\alpha}
X^{\mu}\partial^{\alpha}
X^{\nu}+iB_{\mu\nu}\varepsilon_{\alpha\beta}\partial^{\alpha}X^{\mu}\partial^{\beta}X^{\nu}),\label{4.20}
\end{eqnarray}
we find out that the only difference between two models, is the
antisymmetric matrix $B$, whose non zero components in WZW action is
just $ B_{12}=u $, whereas in Nambu sigma model we have other non
zero components $B_{13}=-a_2,\;B_{23}=a_1$; such that this model is
not conformal. Here, we prove that this model is integrable. For
this purpose, in the following section we use the method presented
by Mohammedi \cite{Moh}.

\section{\bf The integrability of the Nambu-Sigma model on the central extension of the 2D Poincare Lie
group} In this section for self containing of the paper we give a
short review on the integrability of a sigma model, recently
presented by Mohammedi in \cite{Moh}. Consider the following Sigma
model action
\begin{eqnarray}
S_{Ns}=\int_{\Sigma}d^2\sigma
(G_{\mu\nu}\delta_\alpha^\beta+B_{\mu\nu}\varepsilon_{\alpha}^\beta)\partial^{\alpha}x^{\mu}\partial_{\beta}x^{\nu},\,\label{5.1}
\end{eqnarray}
the equations of motion for this model can be written as the
following Lax pair
\begin{eqnarray}
[\partial_0+\alpha_{\mu}(x)\partial_0 x^{\mu}]\psi=0,\label{5.2}
\end{eqnarray}
\begin{eqnarray}
 [{\partial_1}+\beta_{\nu}(x){\partial_1}x^{\nu}]\psi=0,\label{5.3}
\end{eqnarray}
if the matrices $\alpha_{\mu}(x)$ and $\beta_{\nu}(x)$ satisfy the
following relations \cite{Moh}
\begin{eqnarray}
\beta_{\mu}-\alpha_{\mu}=\mu_{\mu},~~~~~\label{5.4}\\
\partial_\mu\beta_\nu-\partial_\nu\alpha_\mu+[\alpha_\mu,\beta_\nu]=\Omega^\lambda_{\mu\nu}\mu_\mu,\label{5.5}
\end{eqnarray}
with
\begin{eqnarray}
\Omega^\tau_{\mu\nu}=\Gamma^\tau_{\mu\nu}-H^\tau_{\mu\nu},\label{5.6}
\end{eqnarray}
 where equation \eqref{5.5} can then be
rewritten by splitting symmetric and anti-symmetric parts as
follows:
\begin{eqnarray}
0=\nabla_{\mu}\mu_{\nu}+\nabla_{\nu}\mu_{\mu}-2\Gamma^{\tau}_{~\mu\nu},~~~~~~~~~\label{5.7}\\
F_{\mu\nu}=-\frac{1}{2}(\nabla_{\mu}\mu_{\nu}-\nabla_{\nu}\mu_{\mu})-H^{\tau}_{~\mu\nu}\mu_{\tau}.\label{5.8}
\end{eqnarray}
In the above formula $\Gamma^\tau_{\mu\nu}$ is the chiristofel
coefficients and
\begin{eqnarray}
H^\tau_{\mu\nu}=\frac{1}{2}g^{\tau\lambda}(\partial_\lambda
b_{\mu\nu}+\partial_\nu b_{\lambda\mu}+\partial\mu
b_{\nu\lambda}),\label{5.9}
\end{eqnarray} such that the field $
F_{\mu\nu} $ and covariant derivative corresponding to the matrices
$\alpha_{\mu}$ are given as follows:
\begin{eqnarray}
F_{\mu\nu}=\partial_\mu\alpha_\nu-\partial_\nu\alpha_\mu+[\alpha_\mu,\alpha_\nu],\nonumber\\
\nabla_\mu X=\partial_\mu X+[\alpha_\mu,X].~~~~~~~~~~~~\label{5.10}
\end{eqnarray}
In this manner, the integrability condition of the sigma model
\eqref{5.1}
 is equivalent to finding the matrices $\alpha_\mu,\mu_\mu$ such that they satisfy \eqref{5.7} and \eqref{5.8}.\\
 Now, we apply this formalism
to the two dimensional Nambu sigma model \eqref{4.18} on the
non-semi-simple Lie algebra \eqref{4.7}. In order to find some
solutions, we proceed by fixing some of these unknowns quantities
with the assumption
\begin{eqnarray}
\alpha_{\mu}=xe^{i}_{\mu}X_{i},\nonumber\\
\mu_{\mu}=ye^{j}_{\mu}X_{j},\label{5.11}
\end{eqnarray}
where the indices of the Lie algebra $i, j, k,...$ have the same
range as those of the target space of the sigma model $\mu,\nu,...$.
We will use the fact that the gauge condition $
A_{\mu}=g^{-1}\partial_{\mu}g=e^{i}_{~\mu}T_{i}$ satisfies the
Bianchi identity $
\partial_{\mu}A_{\nu}-\partial_{\nu}A_{\mu}+[A_{\mu},A_{\nu}]=0$.
As mentioned above, the inverse of the vierbains $e^i_{~\mu}$ are
denoted by $e_i^{~\mu}$ such that we have $
e^{i}_{~\mu}e_{j}^{~\mu}=\delta^{i}_{j}$ and $
e^{i}_{~\mu}e_{i}^{~\nu}=\delta^{\nu}_{\mu}$. We assume that the
quantities $x$ and $y$ are two constant parameters different from
zero. Inserting the expressions of $\alpha$ and $\mu$ in equations
\eqref{5.7} and \eqref{5.8} leads to
\begin{eqnarray}
\Gamma^{\tau}_{\mu\nu}=\frac{1}{2}e_{i}^{~\tau}(\partial_{\mu}e^{i}_{~\nu}+\partial_{\nu}e^{i}_{~\mu}),~\label{5.12}\\
H^{\tau}_{\mu\nu}=\kappa
e^{i}_{~\mu}e^{j}_{~\nu}e_{k}^{~\tau}f^{k}_{ij}\;,~~~~~~~~~\label{5.13}
\end{eqnarray}
with $\kappa=\frac{-1}{y}(x^2-x+xy-\frac{1}{2}y)$. \\
The above Chiristoffel symbols $ \Gamma^{\tau}_{\mu\nu} $ and
torsion $ H^{\tau}_{\mu\nu}$ are those corresponding to the
following metric $g_{\mu\nu}$ and anti-symmetric tensor $b_{\mu\nu}$
\begin{eqnarray}
g_{\mu\nu}=\Omega_{ij}e^{i}_{~\mu}e_{\nu}^{~j},~~~~~~~~~\label{5.14}\\
H_{\mu\nu\tau}=\kappa\Omega_{li}f^{l}_{jk}e^{j}_{~\mu}e^{k}_{~\nu}e_{\tau}^{~i},\label{5.15}
\end{eqnarray}
where $\Omega_{ij}$ is an invertible bilinear form of the
corresponding Lie algebra satisfying
$\Omega_{ij}f^{j}_{kl}+\Omega_{kj}f^{j}_{il}=0$. By solving the
equations \eqref{5.12} and \eqref{5.13} we find that
$\kappa=\frac{3}{2}$ for the Nambu sigma model. To summarise, the
Lax pair construction for the Nambu sigma model represented by the
metric and torsion in \eqref{5.14} and \eqref{5.15} is given by
\begin{eqnarray}
[\partial+x(g^{-1}\partial_{\mu}g)\partial\varphi^{\mu}]\psi=0,\label{5.16}
\end{eqnarray}
\begin{eqnarray}
[\bar{\partial}+\frac{7x}{2x+2}(g^{-1}\partial_{\nu}g)\bar{\partial}\varphi^{\nu}]\psi=0.~~~~~~~\label{5.17}
\end{eqnarray}
\section{\bf Conclusion }
Here, we have presented two dimensional Nambu sigma models on a
manifold in general and on a Lie group as a special case. In
general, these models are not conformal or integrable and one can
obtain conditions under which they have these properties. When the
Nambu structure on a Lie group is of order three, we have obtained
conditions under which this model is equivalent to a WZW model.
Moreover, we have presented an example on the Heisenberg Lie group
for this case. As another example, we have presented the model
constructed by the Nambu structure of order three on the central
extension of the 2D Poincare Lie group, such that this model is
integrable.
\bigskip
\bigskip

\noindent {\bf Acknowledgments:} This research was supported by a
research fund No. 217/d/1639 from Azarbaijan Shahid Madani
University. We would like to thank F. Darabi and Ali Eghbali for
carefully reading the manuscript and useful comments.

\bigskip

\end{document}